\documentclass{aa}

\usepackage{graphicx}
\usepackage{txfonts}
\usepackage[colorlinks=true, allcolors=blue]{hyperref}
\begin{document}

   \title{Plasma line detected by Voyager 1 in the interstellar medium:\\ Tips and traps for quasi-thermal noise spectroscopy}

   \subtitle{ }

   \author{N. Meyer-Vernet
          \inst{1}
          \and
          A. Lecacheux\inst{1}
          \and
          M. Moncuquet\inst{1}
      \and
     K. Issautier\inst{1}
     \and
     W. S. Kurth\inst{2}     
  }

   \institute{LESIA, Observatoire de Paris, PSL Université, CNRS, Sorbonne Université, Université de Paris, 92195 Meudon, France\\
              \email{nicole.meyer@obspm.fr,alain.lecacheux@obspm.fr,michel.moncuquet@obspm.fr,\\karine.issautier@obspm.fr}
              \and
              Dept. of Physics and Astronomy, University of Iowa,  Iowa City, IA 52242, USA\\
              \email{william-kurth@uiowa.edu}
                                                }

   \date{Received 13/07/2023; accepted 25/09/2023}

 
  \abstract{The quasi-thermal motion of plasma particles produces electrostatic fluctuations, whose voltage power spectrum induced on electric antennas reveals plasma properties. In  weakly magnetised plasmas, the main feature of the spectrum is a line at the plasma frequency -- proportional to the square root of the electron density -- whose global shape can reveal the electron temperature, while the fine structure reveals the suprathermal electrons. Since it is based on electrostatic waves,  quasi-thermal noise spectroscopy (QTN) provides in situ measurements. This method has been successfully used for more than four decades in a large variety of heliosphere environments. Very recently, it has been tentatively applied in the very local interstellar medium (VLISM) to interpret the weak line discovered on board Voyager 1  and in the context of  the proposed interstellar probe mission. The present paper shows that the line is still observed in the Voyager Plasma Wave Science data, and concentrates on the main features that distinguish the plasma QTN in the VLISM  from that in the heliosphere. We give several tools to interpret it in this medium and highlight the errors arising when it is interpreted without caution, as has recently been done in several publications. We show recent  solar wind data, which confirm that the electric field of the QTN line in a weakly magnetised stable plasma is not aligned with the local magnetic field. We explain why the amplitude of the  line does not depend on the concentration of suprathermal electrons, and why its observation with a short antenna does not require a kappa electron velocity distribution. Finally, we suggest an origin for the suprathermal electrons producing the QTN and we  summarise the properties of the  VLISM  that could be deduced from an appropriate implementation of QTN spectroscopy on a suitably designed instrument. 
     }
  
\titlerunning{Weak line  in the ISM: quasi-thermal noise with few fast electrons}
\authorrunning{Meyer-Vernet et al.}


   \maketitle
%

\section{Introduction}

A weak continuous line close to the local plasma frequency $f_p$ has been discovered  \citep{ock21} in spectra measured by the Voyager 1 Plasma Wave Science (PWS) instrument \citep{sca77} in the very local interstellar medium (VLISM). Such a continuous line has been detected using long spectral averages and its weakness and stability  suggest that it might possibly be produced by plasma quasi-thermal noise (QTN), despite the small antenna length of the PWS instrument.

The plasma QTN was discovered  in the solar wind by \cite{mey79} with the ISEE-3 radio receiver, which was then the most sensitive receiver ever flown \citep{kno78}. This noise is due to the  electrostatic field produced by the plasma particle quasi-thermal motion \citep{fej69}, detected by a sensitive wave receiver at the ports of an electric antenna. Since this electrostatic field is associated with the plasma velocity distributions \citep{sit67}, in the case of stable distribution functions one can use QTN spectroscopy to reveal plasma properties such as the electron density and temperature \citep{mey89}. This technique has been developed and used in a large variety of media in the heliosphere (\cite{mey98, mey17} and references therein), where the quasi-thermal  noise is routinely observed with wave instruments and represents the long-wavelength limit for radioastronomy measurements from space \citep{mey00}. Because electrostatic waves are heavily damped, the QTN measurements are local, contrary to usual spectroscopy in astronomy. If the plasma is magnetised, then the spectrum has a complex structure including Bernstein waves, from which QTN spectroscopy reveals electron properties \citep{mey93,mon95,sch13}. If the plasma is weakly magnetised, as the interplanetary and interstellar media, then the  QTN spectrum has a much simpler structure, with a line at the local plasma frequency $f_p$ produced by Langmuir waves. The density deduced from this line is recognised as a  gold standard and used as a reference for calibrating other instruments.

\begin{figure*}
        \centering
        \includegraphics[width=17cm]{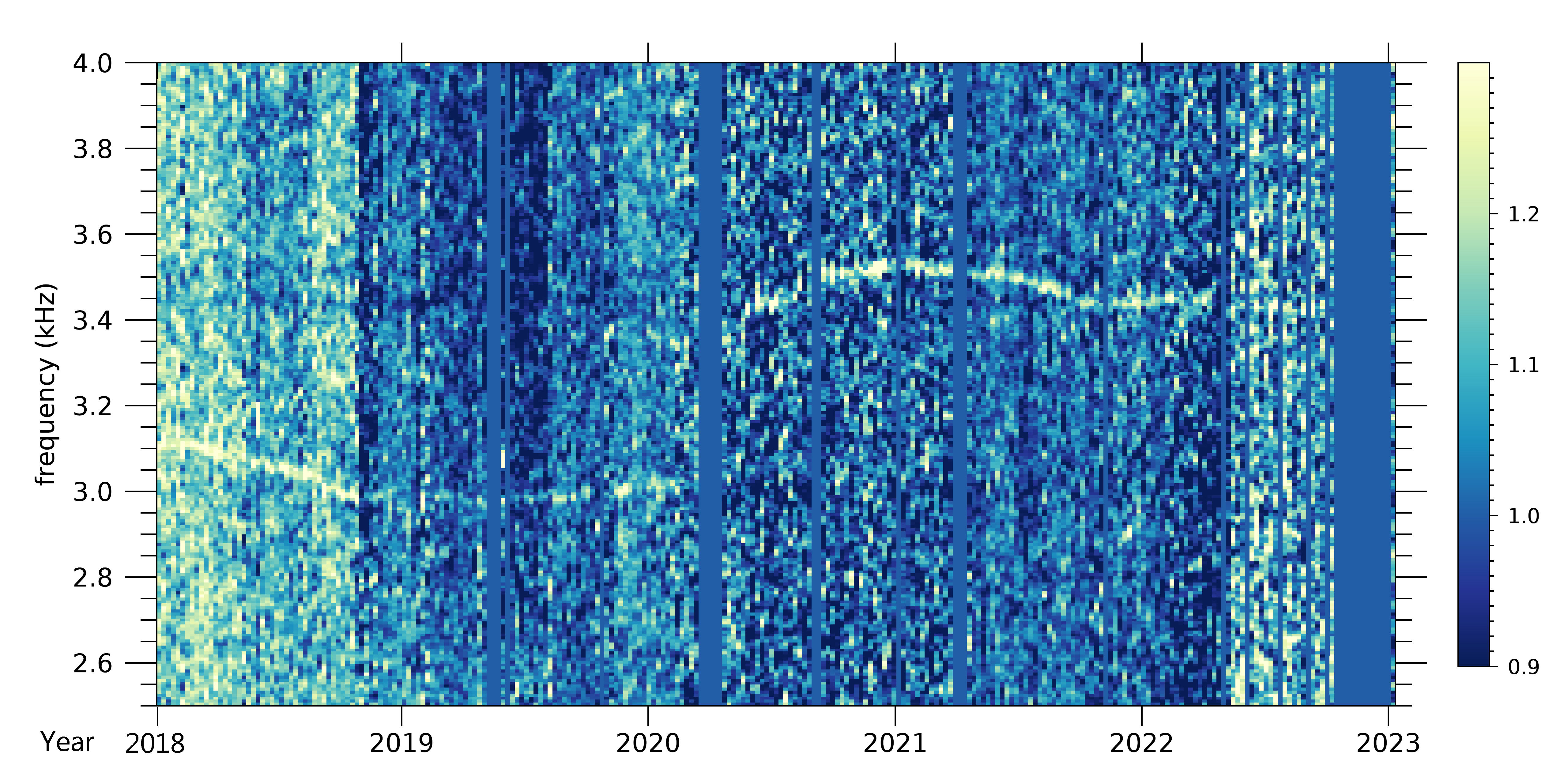}
        \caption{Frequency-time spectrogram showing a portion of the weak continuous line previously published and its continuation from late 2020 until late 2022.  The linear intensity scale is relative to the background level.}
        \label{Fig1}
\end{figure*} 

\citet{gur21} proposed to interpret the $f_p$ line observed by the Voyager 1 PWS instrument in the VLISM as  the QTN associated with an electron velocity distribution composed of the superposition of a cold Maxwellian at 7000$^{\circ}$K and a ten times hotter  kappa  distribution with $\kappa =  1.53$; this proposed hot kappa  distribution represented 50\% of the density and therefore contributed considerably to the pressure. These authors also interpreted the absence of observations of the line before 2016 by arguing that the QTN electric field is oriented along the magnetic field $\textbf{B}$. They suggested  that since the angle $\alpha$ between the Voyager antenna effective direction and $\textbf{B}$ exceeded about 15$^{\circ}$, the resulting weakening of the signal  by the factor $\cos^2 \alpha$ would hinder its observation.

These arguments have been contradicted by \cite{mey22}, who  showed  in a short letter that the stable QTN electrostatic field near $f_p$ in the weakly magnetised VLISM  should  not be aligned with the static magnetic field. Therefore, its orientation could not be responsible for the absence of the line from Voyager 1 PWS measurements taken at locations near the heliopause. In addition to that, \cite{mey22} showed  that a minute quantity of hot electrons with a power-law energy distribution is sufficient to explain the observations, since the amplitude of the line is independent of the proportion of these electrons provided they dominate the distribution at the high speeds producing the line \citep{cha91,mey17}. This property eliminated the need to assume a problematic kappa distribution. Nevertheless, in a recent review paper on the interstellar probe mission, \cite{bra23} repeated the arguments  that the detection of the QTN line at $f_p$ with a modest length antenna requires it to be nearly aligned with the ambient magnetic field and that the observed line can be interpreted with a kappa  distribution with $\kappa =  1.53$.

In this context, the objective of the present study is two-fold: (i) explain how to perform QTN spectroscopy in the VLISM in order to avoid some previous mistakes and  (ii) interpret the observed properties of the $f_p$ line identified on Voyager 1. The paper is structured as follows: section 2 shows that the line continues to be observed in the available recent Voyager 1 PWS data,  with a frequency consistent with \cite{kur23}, and summarises its properties; section 3 presents recent solar wind measurements  illustrating that the electric field of the stable QTN line in a weakly magnetised plasma is not aligned with the local magnetic field; section 4  discusses the main differences between QTN spectroscopy in the heliosphere and in the VLISM, the traps to be avoided, and some useful tips; section 5 suggests several explanations for the absence of the line close to the heliosheath and proposes an origin for the suprathermal electrons producing the observed $f_p$ line; and finally, we summarise the properties of the VLISM  that could be derived  with adequate instrumentation and implementation of QTN spectroscopy.

\section{The Voyager QTN line
}   
 
Figure 1 shows the weak continuous line near the local plasma frequency $f_p$ measured from the Voyager 1 PWS wideband data in the VLISM from early 2018 to late 2022. The line observed before late 2020 was published and discussed previously \citep{ock21,bur21,gur21,ric22,mey22,bra23}. The two-year line continuation in 2021-2022 is consistent with the observations shown by \cite{kur23}.

The spectrogram is built from fast Fourier transforms (FFT) of Voyager 1 PWS waveform data. These  waveform data were designed to fit within an imaging subsystem (ISS) image frame. This frame is made of 800 lines, each filled up with 1600 4-bit waveform data at the 28.8 kHz sampling frequency, and is written in 48 seconds (including small data gaps between individual lines); for details, readers can refer to \cite{kur23}. All spectra used in the present analysis are the average of single FFT power spectra computed from each individual line. 
Because of a mismatch between the Deep Space Network and Voyager playback capabilities, only one out of every five of those 800 lines can be transmitted to Earth, making any enhancement of the spectral resolution by FFTing consecutive lines impossible. The best available spectral resolution of PWS wideband data is therefore limited to 28800/1600 = 18 Hz -- a limit that might be easily overcome in any future dedicated instrument. The noise equivalent bandwidth (NEBW) of the measurement  is increased from 18 Hz to about 24 Hz because of the apodisation (Hamming window).

\begin{figure}
        \centering
        \includegraphics[width=9cm]{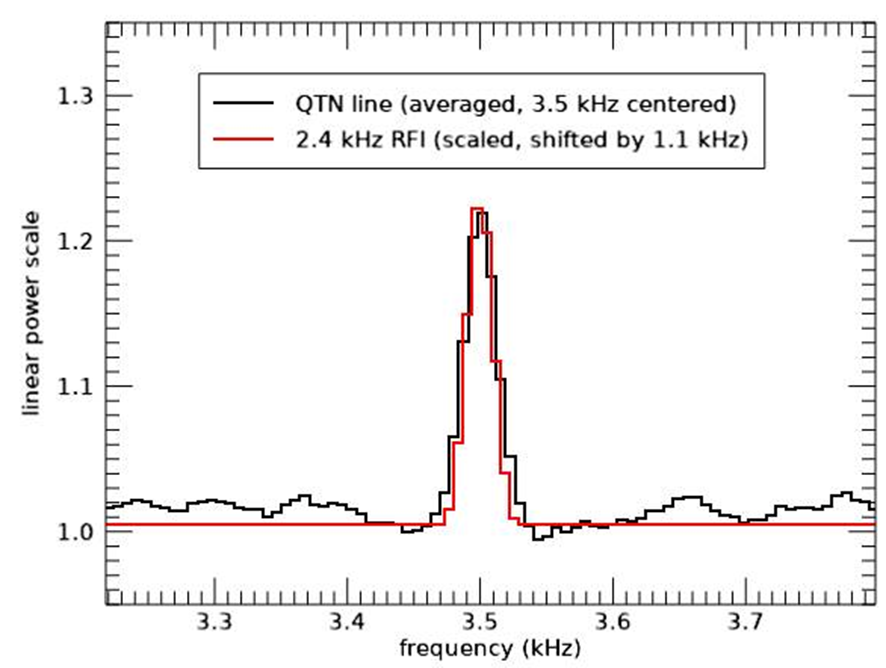}
        \caption{Observed $f_p$ line (in black, linear scale) superimposed on the  supply interference line (in red) scaled in amplitude and shifted in frequency. The $f_p$ line profile was obtained by averaging 45,300 good-quality spectra which were shifted to a common central frequency, fitted to a Gaussian profile, and acquired within the period shown in Figure 1. The total power exceeds the background by about 20\%, which means that the contribution of the line amounts to roughly 20\% of the background.
        }
        \label{raie1822}
\end{figure}

The measured line width is nearly 27 Hz, which is not significantly larger than the instrumental NEBW given the uncertainties. It follows that  the line is not resolved, having a measured width close to the frequency resolution. This absence of resolution can be checked by comparing the observed line to the 2.4 kHz interference line, whose measured width appears similar despite its presumably quasi-Dirac shape (Figure \ref{raie1822}). So the intrinsic width of the $f_p$ line is expected to be smaller than (or equal to) 24 Hz. The figures were obtained using averages over a part (about 10 seconds) of the recorded 48 second data snapshots, spaced by 2.2 days or 7 days, depending on the available telemetry, without any detectable change in line width over the instrumental value.

With a 7 day temporal resolution, the only notable event is a strong density increase of  roughly 36\% in  2020, with a rapid variation  around May 2020 associated with a similar increase in  magnetic field strength \citep{bur21}. This event took place unfortunately just after a data gap of about one month.

\begin{figure}
        \centering
        \includegraphics[width=8cm]{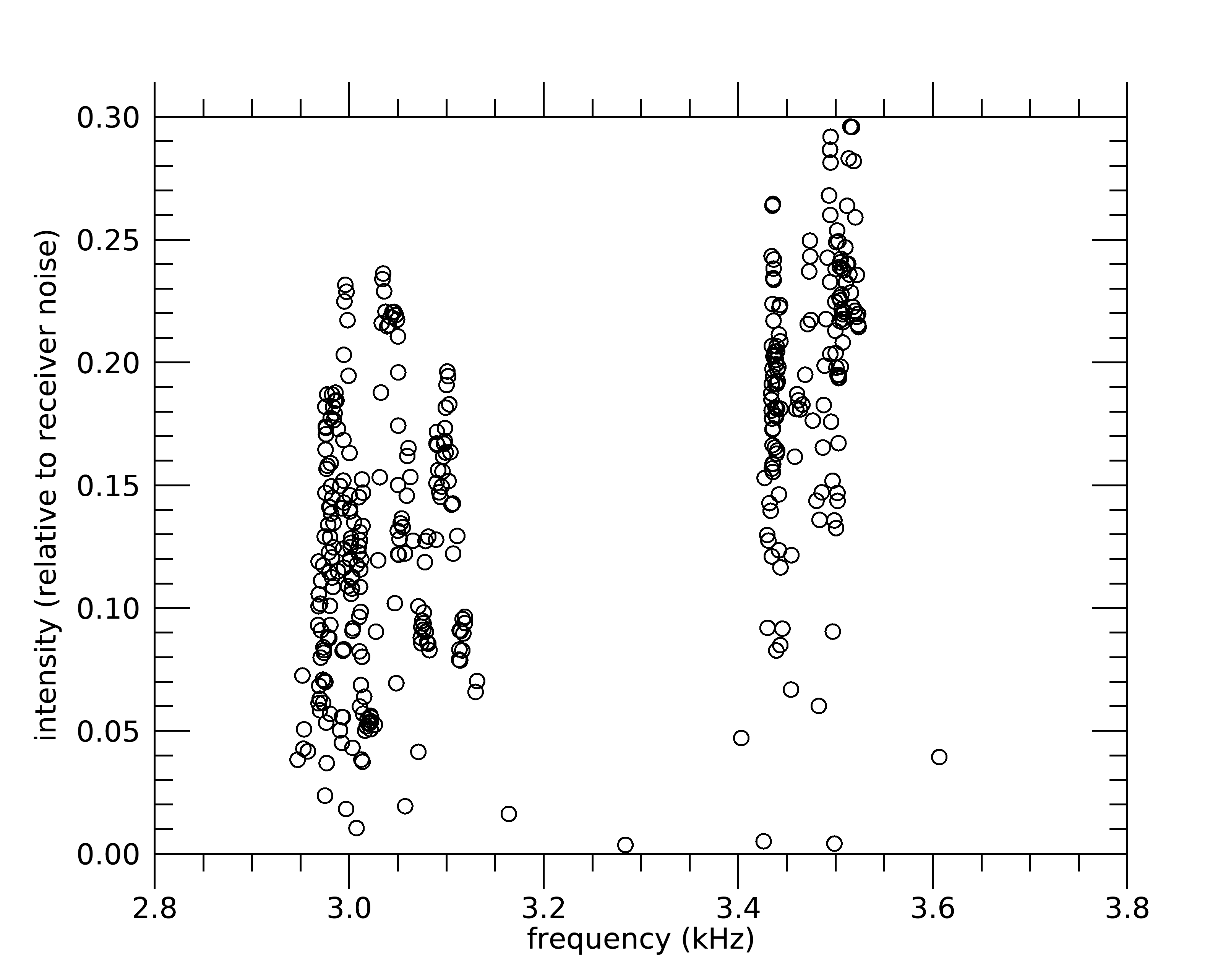}
        \caption{Intensity of the line relative to the background as a function of its frequency, for the spectra acquired within the period shown in Figure 1. The average power of the line of $14 \pm 6 $ \% near 3 kHz  increases to  $20 \pm 4 $ \% near 3.5 kHz, which is in  agreement with Eq.\ref{peakfin}.
                                }
        \label{amplitude}
\end{figure}

Over the 5 years shown in Figure 1 (from early 2018 to late 2022),  the  contribution of the QTN line amounts to about 20\% of the  background on average. This contribution is roughly two times higher than its mean value from September 2017 to late 2020 \citep{ock21, mey22}, suggesting that the amplitude of the line increased, as shown in Figure \ref{amplitude}, where the line intensity is displayed as a function of its  frequency. This increase, associated with the   increase in density ($n \propto f_p^2$), can be entirely attributed  to the variation in the line intensity in proportion of $f_p^{2.5}$ in Equation \ref{peakfin} (see section 5). However, we note that the use of the automatic gain controlled (AGC) receiver and no telemetered information on the gain as well as in some cases extreme noise due to a low signal-to-noise ratio in the telemetry link makes this comparison somewhat uncertain.

\section{QTN $f_p$ line and magnetic field direction
}

    \begin{figure*}
        \centering
        \includegraphics[width=13cm]{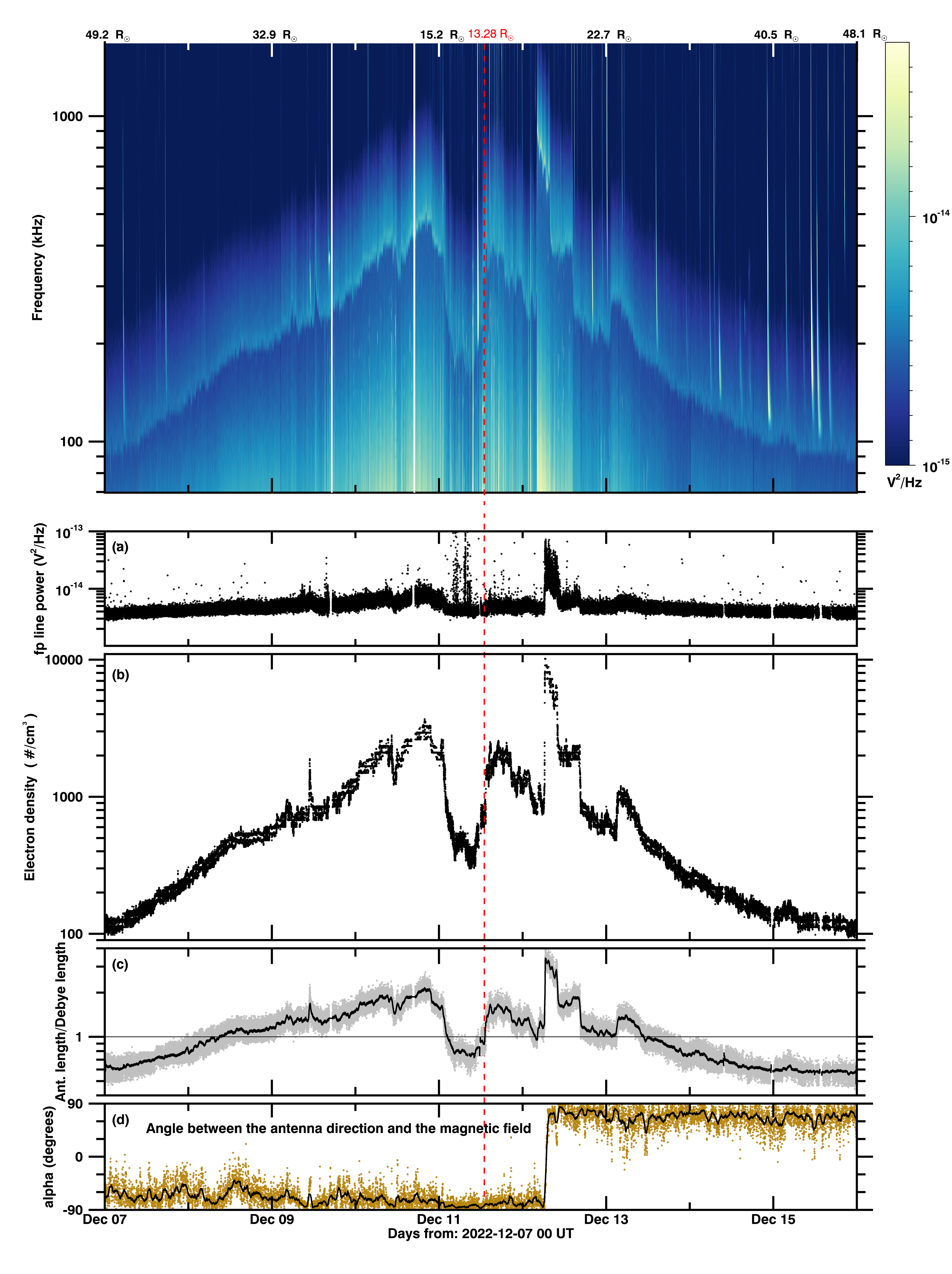}
        \caption{Top panel. Spectrogram acquired during the 14th PSP solar perihelion (closest solar distance 13.28 solar radii) with the FIELDS antenna, showing the plasma QTN, on which the $f_p$ line clearly emerges (cyan line varying between 90 kHz and 1 MHz). The heliocentric distance in solar radii is indicated at the top. Bottom panel. (a) Spectral power at the  peak, (b) total electron density, (c) ratio of the antenna length to the plasma Debye length, and (d) angle  between the antenna direction and the magnetic field, showing  a 180$^{\circ}$ variation at the heliospheric current sheet crossing (courtesy of FIELDS MAG). The superimposed black lines in panels (c) and (d) are one-hour rolling averages. The vertical dotted red line indicates the time of perihelion.}
        \label{FigPSP}
   \end{figure*}

As we noted in section 1, \cite{gur21} and \cite{bra23} have argued that the electric field of the QTN $f_p$ line is aligned with the ambient static magnetic field, in order  to  explain why  the line was not detected on Voyager near the heliopause.

This argument is not expected to be correct in weakly magnetised stable plasmas, where the electron gyrofrequency $f_B = eB/(2 \pi m)$ is negligible with respect to the plasma frequency. In the interstellar medium where the line is detected, $f_B/f_p \sim 4 \times 10^{-3}$ (e.g. \cite{bur21}). It follows that, with the values of the wave number $k$ contributing to the line, the $f_B$ term is negligible  in the equation of the generalised Langmuir mode  \citep{wil00} $f^2(k, \theta) = f_p^2+f_B^2 \sin^2 \theta + (k v_{\mathrm{th}}/2 \pi)^2 $, with    $\theta$ being the angle between  $\textbf{B}$ and the longitudinal electric field   \citep{mey22}. 
No variation in the QTN with the angle  between the antenna direction and the magnetic field has ever been observed in weakly  stable plasmas  \citep{mey20} during four decades of QTN observations, except a small variation due to the anisotropy of the electron temperature \citep{mey94}.

Figure \ref{FigPSP}  shows a counter example of the claimed necessity for the antenna direction to be aligned with the local static magnetic field for measuring the QTN $f_p$ line in weakly magnetised  plasmas. The top panel shows a spectrogram  measured by the FIELDS instrument \citep{bal16,pul17} in the solar wind during the 14th  perihelion of Parker Solar Probe (PSP). In the bottom part of Fig. \ref{FigPSP}, panel (a) shows the spectral power at the $f_p$ peak, which is used on PSP to estimate the temperature of the suprathermal  component of the electron velocity distribution \citep{mon20}, panel (b) shows the  electron density deduced from the spectra, panel (c) shows  the ratio of the antenna length to  the Debye length deduced as described in the paper cited above, and panel (d) shows the angle  $\alpha$ between the antenna direction and the static magnetic field. With the ratio $f_B/f_p \sim (1-3) \times 10^{-2}$, the angle  between the antenna direction and the magnetic field varies between 45 and 90$^{\circ}$  without affecting the amplitude of the QTN line, with a ratio between the antenna length $L$ and the Debye length $L_D$ between 0.6 and 2.

\section{Tips and traps for QTN spectroscopy in the interstellar medium
}

QTN measurements in the VLISM differ from those made currently in the heliosphere \citep{mon20} by several major aspects. First, the spatial scales are generally much larger  in the VLISM (e.g. \cite{fra22},  \cite{ric23}). So, the properties of the interstellar medium are  generally much more constant in  space and time (as seen from a spacecraft) than in the heliosphere. Furthermore, the high-frequency compressible  turbulence has a much smaller amplitude (\cite{ock21} and references therein). The resulting quasi-constancy of the electron  density enables the spectra to be averaged for much longer times, so very weak features can be detected. For example, the Voyager interstellar line shown in Figs \ref{Fig1} and  \ref{raie1822} was detected by sampling the data over  times that  could be separated by one week,   whereas the solar wind line shown in Figure \ref{FigPSP}  was  measured  with an acquisition  time  $\simeq 2$ s by the Low Frequency Receiver of the FIELDS instrument on PSP    \citep{pul17}. The QTN noise is indeed rarely integrated for more than a few seconds in the interplanetary medium because of the short-wavelength density fluctuations \citep{cel87}, which widen the $f_p$ line \citep{cha91}. 
This property enables one to detect the QTN line far below the instrumental noise of Voyager PWS, since the averaging increases the level of the signal to be detected compared to the fluctuations of the instrumental noise. The  line measured in the interstellar medium does indeed have  an average power of 10-20 \% of the receiver noise  \citep{ock21, mey22}. In contrast,  the power spectral density in the PSP line shown in Figure \ref{FigPSP} is of the order of magnitude of $ 10^{-14}$ V$^2$/Hz, which is higher than the instrumental noise (about $2.2 \times 10^{-17}$ V$^2$/Hz) by more than two orders of magnitude.

Second,   the  electron density is very small in the interstellar medium, which has two important consequences. The particle-free paths are very large. The  Debye length is relatively large, exceeding the equivalent length of the Voyager antenna $L\simeq 7.1$ m \citep{gur21}. It is well known that in that case, the  QTN $f_p$ line is minute in a Maxwellian plasma and still difficult to detect in the presence of a hot suprathermal Maxwellian component \citep{mey89}. The Debye length
\begin{equation}
L_D = [\epsilon_0 k_B T_{-2}/(ne^2)]^{1/2} \label{LD}
,\end{equation}
where
\begin{equation}
k_B T_{-2}/m = 1/\langle  v^{-2} \rangle  
\;\; \text{with} \;\; \langle v^{-2} \rangle = \int  d^3 v\; v^{-2}  f(v) \label{T-2}
,\end{equation}
with $f(v)$  being the electron 3D velocity distribution normalised to the electron density, is determined by the coldest electrons. Hence, when the distribution is composed of the superposition of a cold Maxwellian and a small proportion of suprathermal electrons, $L_D$ is determined by the temperature of the cold Maxwellian.

Third, although  the shot noise is  often a nuisance in the interplanetary medium, requiring the antennas to be very thin \citep{mey17}, the shot noise is expected to be very small  in the interstellar medium because the photoelectron emission by the antenna is much smaller than  the plasma electron current. This produces a negative antenna potential of a few times the electron energy, which strongly reduces the flux of incoming plasma  electrons  \citep{whi81}, and  therefore the shot noise \citep{mey17}.
 
In general, a few theoretical properties of the QTN enable a simple measurement in  weakly  magnetised plasmas \citep{mey89,mey17}, as the interplanetary and interstellar media. First, the plasma frequency reveals the electron density, provided the line emerges from the rest of the spectrum. And since the Langmuir wavelength tends to infinity at $f_p$, the QTN measurement is equivalent to a detector of a large cross-section and is relatively immune to spacecraft perturbations \citep{mey98}.

Second, below the plasma frequency, the electron QTN  spectrum is determined by electrons crossing a Debye length around the antenna. Each such electron induces a potential pulse of duration $\sim 1/(2\pi f_p )$, producing a plateau below $f_p$ with an amplitude mainly depending on the cold component of the electron distribution. Although the level of this plateau can be calculated numerically  \citep{mey89, mey17}, an approximate measurement can be easily made via the following analytic formulas:\\
 $L/L_D\ll1$: $ V^2 \simeq [(2k_B T m)^{1/2}/(3 \pi ^{3/2} \epsilon_0) ]
(L/L_D)^2 [1+\ln(L_D/L)] \simeq 3.4 \times 10^{-17} T^{1/2} (L/L_D)^2 [1+\ln(L_D/L)]$, \\
 $2<L/L_D<7$:  $ V^2 \simeq (k_B T m)^{1/2}/(\pi ^{2} \epsilon_0) \simeq 4.1 \times 10^{-17} T^{1/2} $, and  \\
$L/L_D\gg1$: $ V^2 \simeq (\pi /2)^{1/2} k_B T/( \epsilon_0 \omega_p L) \simeq  3.5 \times 10^{-14} T/(n^{1/2}L) $.\\
Here,  $T$ is the electron temperature, $m$ is the electron mass, and $\omega_p = 2 \pi f_p$.
When the electron velocity distribution is the superposition of a cold Maxwellian and a hot dilute halo, the temperature in these formulas is roughly that of the cold component, as for the expression of the Debye length. For more complex distributions, detailed values are given by \cite{mey17}. The above expressions neglect the contribution of the ions  \citep{iss99}.

Third, the high-frequency spectrum for $(f/f_p)(L/L_D) \gg1$ is proportional to the electron total pressure: $V^2 \simeq f_p^2 k_B T/(\pi \epsilon_0 L f^3)$.

We now consider the $f_p$ peak. The Langmuir wave number at frequency $f=f_p+\Delta f$ with $\Delta f/f_p \ll 1$ is 
$k_L \simeq (\omega_p/v_{\mathrm{th}}) [2 \Delta f/f_p]^{1/2} $, 
where $v_{\mathrm{th}}$ is the electron root-mean-square speed defined as
\begin{equation}
v_{\mathrm{th}}^2 =  \langle v^2 \rangle = \int_0^\infty d^3v \; v^2 f(v) = 3 k_B T/m  \label{vth}
,\end{equation}
which also defines the temperature for velocity distributions that are not necessarily Maxwellian. The speed of the electrons producing the QTN at  $f= f_p+\Delta f$ (with $\Delta f/f_p \ll 1$) is therefore
\begin{equation}
v_{\mathrm{ph}} \simeq  \omega_p/k_L = v_{\mathrm{th}} [ f_p/2\Delta f]^{1/2}   \label{vph}
.\end{equation}
From equations (\ref{vth}) and (\ref{vph}) with the assumed temperature $T\simeq 7,000$ $^{\circ}$K \citep{mcc15}, we deduce that the QTN at frequencies in the range $f_p < f < f_p+\delta f$, where  $f_p \simeq 3.5$ kHz and $\delta f = 12 $ Hz is the maximum half-width of the line (section 2), is  produced by electrons with a speed faster than $v_{ph}\simeq 6.8 \times 10^6$ m/s, which corresponds to energies above about 100 eV.

The QTN  at frequency $ f_p+\Delta f$, where $\Delta f/f_p \ll1$, measured with an antenna of length $ L$, is given by \citep{mey17} 
\begin{equation}
V_f^2 \simeq \frac{8 m v_{ph} F(\omega_pL/v_{ph})}{\pi \epsilon_0  v_{th}^2} \left[\frac{ \int _{v_{ph}}^{\infty}  d v \; v \; f(v)}{ f(v_{ph})} \right] \label{peak}
,\end{equation}
where $v_{\mathrm{ph}} $ is given by (\ref{vph}) and  $F(x)$ is the antenna response. Since  $\omega_p L /v_{\mathrm{ph}} \ll 1$, we have 
$F(x) \simeq x^2/24 $ \citep{mey89}.

The width  of the line observed on Voyager is too small to be produced by a hot suprathermal Maxwellian \citep{mey22}, contrary to the QTN line generally observed in heliospheric plasmas. Such a thin line can be produced by a minute amount of hot electrons with a power-law energy distribution.  Superimposing on a Maxwellian at temperature $T$, such  a distribution $f_h (v) \propto 1/v^s$ with $s > 2$ at energies exceeding $\simeq 100$ eV yields a distribution whose thermal speed  $v_{th}$ is determined by the Maxwellian and whose value at  $v \geq v_{ph}$ nearly equals  $f_h (v)$. Therefore, the square bracket in (\ref{peak}) is determined by $f_h (v)$ and independent of the concentration of these electrons, and the QTN at frequency $f_p+\Delta f$ (with $\Delta f/f_p \ll 1$) is given by
\begin{equation}
V_f^2 =  \frac{2^{3/2}  \pi  m f_p^2 L^2}{3(s-2) \epsilon_0 v_{th}}  \left(\frac{f_p}{\Delta f}\right)^{1/2} \label{peakfin}
.\end{equation}

The power associated with frequencies closer to $f_p$ than $\Delta f$  should be larger than this value in the absence of widening of the line by density fluctuations, but it cannot be measured at frequencies closer to $f_p $ than the frequency resolution, characterised by the instrumental NEBW. Hence the measured line width should be of the order of the frequency resolution, as observed. (We note   that if the instrumental NEBW were much smaller,  the  speed of the electrons producing the peak would become relativistic.)

Approximating the peak level by the average of (\ref{peakfin}) between $f_p $ and $f_p+\delta f/2$, corresponding to the half-width, and 
substituting $f_p \simeq 3.5$ kHz, $L \simeq 10/2^{1/2} \simeq 7.1$ m, $T\simeq 7,000$ $^{\circ}$K, and $s = 5$, we obtain $V_f^2 \simeq  4 \times 10^{-15}$ V$^2$/Hz. Adding the remaining QTN noise of the order of the  plateau  $\simeq 10^{-15}$ V$^2$/Hz from the expressions written above, we get $V_f^2 \simeq 5 \times 10^{-15}$ V$^2$/Hz. With the published receiver noise of about  $10^{-13}$ V$^2$/Hz \citep{kur79, gur21}, the measured line shown in Figure \ref{raie1822} has an average intrinsic  level  of the order of $ 2 \times 10^{-14}$ V$^2$/Hz, which is a few times larger than our theoretical estimate. However, the published   receiver noise is based on the spectral density noise threshold of the spectrum analyser channels, whereas for the present work we used  the waveform receiver, which could detect line emissions below this level with sufficient signal averaging, so the values of the theoretical and observed line levels are marginally compatible. We note that (\ref{peakfin}) shows that the intensity  of the line would  decrease if $T$ or the instrumental NEBW were larger. We shall return to this point in section 5.

We now evoke some traps into which one may fall when trying to apply QTN spectroscopy in the VLISM. The QTN calculations published by \cite{gur21} and  \cite{bra23} constitute interesting examples of these traps. We note that in Figs 4 and 5  of the former paper, the plotted shot noise, which represents the main noise contribution below $f_p$, is  too large by factors of 10 and 20   for $f<f_p$, respectively, even with a positive antenna potential as in the model by \cite{mey89}  cited in the caption of these figures. These errors may have arisen in particular because the antenna impedance  should be calculated correctly in this frequency range \citep{zou09}. A further little known trap is that the slope of the shot noise changes for $f>f_p$, with a  decrease much faster than $1/f^2$, because the rise time of the voltage pulses producing the shot noise is roughly the time for an electron to travel a Debye length. It follows that the squared Fourier transform decreases much more steeply than $1/f^2$ for $f>f_p$ \citep{mey85}. Furthermore, with the expected small photoelectron emission from the antenna in the VLISM, the shot noise should become much smaller because of the expected negative antenna potential. In addition, the ion QTN \citep{iss99} is not negligible below $f_p$ with the parameters considered in the figure by \cite{gur21} reproduced by \cite{bra23}.

We now consider the proposed interpretation of the $f_p$ line by the superposition in equal proportions of a Maxwellian and a ten times hotter kappa distribution with  $\kappa = 1.53$  (Figure 7 by \cite{gur21}, reproduced in Figure 14 by \cite{bra23}). This highly publicised figure  exhibits further traps.

As we already noted, the $f_p$ peak is determined by the suprathermal electrons, independently of their concentration. More precisely, the noise at frequencies between $f_p$ and $ f_p +\Delta f$ is determined by the shape of the electron distribution at speeds exceeding the value given in Eq.(\ref{vph}). At such speeds, the  contribution of the Maxwellian has considerably decreased, so the distribution assumed in these papers reduces to the kappa function, which itself reduces to a power law  $ \propto 1/v^s$ with $s = 2 \times (\kappa  +1) \simeq 5$. Therefore, the QTN at frequency $f = f_p +\Delta f$  with the distribution assumed in these papers should be roughly equal to the value given in (\ref{peakfin}). However, as we already noted, the smaller the value of $\Delta f$, the higner the intensity, and the theoretical QTN should not be plotted closer  to $f_p$ than the frequency resolution in order to compare it to the data. We note, finally, that such a figure showing a QTN peak level close to the receiver noise contradicts the observations, which show a much smaller value.

The caption of this figure is  interesting, too,  since it reveals a frequent  misunderstanding of the QTN. The caption attributes the large level of the  peak to the smallness of the Debye length of the kappa distribution. The origin of such a misunderstanding is that increasing the thermal speed $v_{th}$ indeed decreases the peak level and that the Debye length is proportional to  $v_{th}$ if the plasma is Maxwellian. However, the small Debye length of  a kappa distribution with a value of kappa close to 1.5 is produced by the small value of $T_{-2}$ given by (\ref{T-2}), which has no effect on the peak, independently of the problems raised by such a distribution \citep{mey22}.

\section{Discussion and conclusion}

If the absence of the stable $f_p$ line close to the heliopause does not come from the  angle between the Voyager antenna direction and the magnetic field, we must find an alternative explanation. 
\cite{mey22} suggested that the density fluctuations, which do not prevent the detection of the  line  farther out, may increase close to the heliopause, where compressive fluctuations in the heliosheath are transmitted  \citep{bur15, bur18}, without reaching large distances farther out \citep{zan19}. This may broaden the peak and therefore decrease its amplitude. This suggestion should be studied in detail, which is outside the scope of the present paper. Other possible explanations should be examined as well. First, the  electron density is much smaller close to the heliopause, especially before 2015 \citep{kur23}, which would decrease the intensity of the line  according to equation (\ref{peakfin}). Second, this equation shows that the line intensity should decrease close to the heliopause  if the electron  temperature increases   (e.g. \cite{fra21} and references therein) since the peak varies as the inverse of the thermal speed. In particular, if $T \simeq 30,000$ $^{\circ}$K or more as the ion temperature measured by Voyager 2 PLS immediately outside the heliopause, albeit with currents  close to the instrument threshold  \citep{ric19}, the intensity of the line would decrease by a factor of two or more, as well as the QTN plateau. A third possibility is a variation in the energy spectrum of the suprathermal electrons and of their minimal energy which determines the intrinsic width of the line via equation (\ref{vph}).

On the other hand, as we noted in section 2, the increase in line intensity in 2020 from 14\% to 20\% of the background  when the frequency of the $f_p$ line increases from roughly 3 kHz to 3.5 kHz (Figure \ref{amplitude}) is entirely attributable to this increase in the value of $f_p$ since equation \ref{peakfin} shows that the intensity of the line should be proportional to $f_p^{2.5}$.

An important question is: what is the origin of the  suprathermal electrons of energy exceeding $\sim 100 $ eV assumed in the present paper in order to produce the QTN line? This energy is of the same order of magnitude as that of the electron beams  suggested to produce the instability exciting the  plasma oscillation events detected close to the heliopause \citep{gur21}.

We first note that the presence of suprathermal electrons near  100 eV  is not surprising  since with an ambient electron density of the order of $0.15$ cm$^{-3}$, their Coulomb-free path is much larger than the distance from the heliopause and other density gradients  that might produce them. 

We suggest below an original explanation for these suprathermal electrons: the presence of density gradients. In order to enforce plasma  quasi-neutrality, density gradients produce ambipolar electric fields, of order of magnitude $E$ given by the electron momentum equation, which can be approximated by  $eE \simeq k_B T/H$, where the scale height $H$ is roughly given by $H^{-1} \simeq n^{-1}(dn/dx)$. Here, $dn/dx$ is the space derivative of the density and  the temperature gradient is neglected compared to that of the density in the electron momentum equation. If the electric field exceeds the Dreicer field $E_D$ given by $eE_D \simeq 2k_BT/l_f$ \cite{dre59,dre60}, with $l_f$ being the mean-free path of thermal electrons, electrons with an energy exceeding the thermal energy times $3E_D/E$ may undergo runaway, yielding a non-thermal velocity distribution. \cite{scu19,scu23} suggested steady runaway as the origin of the ubiquitous  suprathermal electrons in the  solar wind and possibly other astrophysical contexts. Such a production of suprathermal electrons above about $\sim 100$ times the thermal energy would thus require density  gradients of scale height $H \sim 100 \; l_f/6$.  With a Coulomb-free path of thermal electrons $l_f \simeq 0.15 $ AU in the LISM measured from Voyager 1's  available data, the required scale height would be a few  AU if such a process acts in a steady way. This is similar to the scale height reported by \citet{gur13} and \citet{kur23}.

Finally, it is important to note that the available data make the analysis difficult. In addition to telemetry errors, observational gaps, and other problems, the very high instrumental noise requires long spectral averages to detect the  line. Furthermore, the intrinsic power of the signal is unknown since the level of the automatic gain control was not telemetered, so the signal can only be deduced relative to the very large instrumental noise. With a modern sensitive instrument and antennas of 50 m length (see Interstellar Probe Concept Study Report at \url{interstellarprobe.
jhuapl.edu}), the QTN plateau enabling a simple measurement of the thermal electrons could be easily measured, as well as the  plasma frequency peak, even in the absence of a significant suprathermal component. Measuring a suprathermal component with a power-law distribution for electrons of energy exceeding 100 eV as considered here requires  an instrumental relative NEBW of the order of $3 \times 10^{-3}$.

Further analysis should be performed to check the proposed mechanisms using future data from Voyager, which can bring new perspectives for the Interstellar Probe project.

\begin{acknowledgements}  
W.S.K. acknowledges support by NASA through Contract 1622510 with the Jet Propulsion Laboratory and the use of the Space Physics Data Repository at the University of Iowa supported by the Roy J. Carver Charitable Trust.
\end{acknowledgements}

%
%

\end{document}